# Numerical investigation of the effect of high voltage frequency on the density of RONS species in the air atmospheric pressure gas discharge

**Fariborz Momtazzadeh [1], Farshad Sohbatzadeh [1,2], Hamed Soltani Ahmadi [1,2], Ramin Mehrabifard[3]**

[1]Department of Atomic and Molecular Physics, Faculty of Basic Sciences, University of Mazandaran, Babolsar, 47416-95447, Mazandaran, Iran

[2]Plasma Technology Research Core, Faculty of Science, University of Mazandaran, Babolsar, Iran

[3]Division of Environmental Physics, Faculty of Mathematics, Physics and Informatics, Comenius University, Mlynska dolina, 842 48 Bratislava, Slovakia

*Corresponding e-mail: hamedsoltani1991@gmail.com

## Abstract

In the last few decades, studies in various fields of plasma technology have expanded and its application in different processes has increased. Therefore, the achievement of a desirable and practical plasma with specific characteristics is of particular importance. The frequency of the applied voltage is one of the important factors that play a role in the physical and chemical characteristics. In this research, changes in the density of active species produced in an electrical discharge using a dielectric barrier and air working gas have been investigated, from a frequency of 500 Hz to 500 kHz, and by applying a constant voltage of 2 kV, have been investigated. For this purpose, 87 different reactions with specific collision cross-sections were defined in COMSOL Multiphysics. Other parameters including current-voltage waveform, electric field, species density including NO, OH, $O_2^+$, $O_3$, etc. were evaluated. The results show that under completely identical conditions, the electron temperature distribution changes with increasing applied frequency, and the density of reactive oxygen and nitrogen species (RONS) like (NO, OH, $O_3$,..) decreases, but $O_2^+$ shows an increasing trend. It should be noted that the simulation results are in good agreement with previous experimental and simulation reports. These results offer valuable insights into optimizing plasma parameters for different applications, potentially resulting in better treatment outcomes across a range of therapeutic domains.

**Keywords:** Cold Atmospheric Pressure Plasma, Dielectric Barrier Discharge, RONS, High Voltage Frequency, COMSOL Multiphysics, Gas discharge

## I. INTRODUCTION

Atmospheric pressure cold plasma, as an innovative technology, has diverse applications in various fields, surface treatment [1], water decontamination [2], and many applications in medicine [3], including killing bacteria [4], cancer cell treatment [5]–[7] and so on. Plasma is a quasi-neutral gas composed of positive and negative ions, free radicals, UV radiation, and other particles, with approximately equal charge density [8], [9]. Since the ionization process typically starts at a specific temperature, usually several thousand Kelvins, plasma is often referred to as the fourth state of matter [10]. It is important to note that not all ionized gases can be classified as plasma, as specific conditions must be met for a state to be defined as plasma.

One method of plasma generation involves the electrical discharge of various gases using a dielectric barrier, which is achieved by applying the appropriate voltage in different frequency ranges (with sufficient injected power) [11].

The electron is freed first from the atoms and molecules by natural background radiation, cosmic rays, or thermionic emission, and then the electron gets accelerated by the electric field. Accelerated electrons produce ionized atoms and molecules. The plasma generated by gas discharge possesses unique characteristics that determine its applications in different fields. Therefore, investigating the physical and chemical parameters of plasma under various conditions is crucial to design an optimized plasma reactor tailored to specific applications [12].

Simulation is a method wherein a computer model of a real or hypothetical system is developed to analyze its behavior under varying situations [13]. These models can simulate physical, chemical, biological, or even social phenomena, and are used for predicting outcomes, optimizing processes, and training purposes [14]. Simulations are widely used in engineering, medicine, economics, and social sciences [15]. This method allows users to analyze desired results without the need for real and costly experiments [16].

COMSOL is one of the most advanced and comprehensive simulation tools in the fields of science and engineering. This software, as a comprehensive and advanced package to simulate complex systems, is capable of solving differential equations of nonlinear systems using partial derivatives using the Finite Element Method (FEM) in various spatial dimensions, including zero, one, two, and three dimensions, accurately and effectively [17].

In recent years, extensive studies have been conducted on the simulation of dielectric barrier discharge (DBD) [18]–[20]. For example, in 2002, Golubovskii and colleagues performed a numerical simulation on helium plasma and calculated various plasma parameters under constant conditions [14]. In 2009, Petrovic et al. conducted a two-dimensional simulation of cylindrical DBD with helium gas mixed with nitrogen impurities [15]. In 2013, Rokeya and colleagues performed a one-dimensional simulation for DBD with helium gas, analyzing the effects of dielectric constant and electrode spacing [16]. In 2017, Gadkari and colleagues simulated a coaxial cylindrical DBD plasma in pure helium gas using a two-dimensional fluid model in COMSOL [17]. Furthermore, in previous study in our group a one-dimensional simulation in COMSOL performed and useful plasma parameters calculated [18]. We also investigated the effect of humidity on helium discharge and its effect on RONS production [21].

This study numerically examines the effect of changing the applied high-voltage frequency in atmospheric pressure electrical discharge on the density of active oxygen and nitrogen species in air. Determining useful parameter values experimentally is not cost-effective; thus, simulating an air gas discharge system using a dielectric barrier within a frequency range of 500 Hz to 500 kHz is necessary. This simulation examines parameters such as species density, voltage-current characteristics, electric field distribution, electron temperature distribution, average species energy distribution, etc. The specific frequencies analyzed in this simulation are 500 Hz, 50 kHz, and 500 kHz. A significant observation is the decline of some species with increasing frequency. The goal of our study is to highlight the dynamic features of the ignition phase and examine how discharge characteristics vary with frequency in this early stage.

## II. SIMULATION PROCEDURE

In this section, the simulation of dielectric barrier discharge in the presence of a nitrogen gas is described. This simulation allows us to analyze and examine key variables such as species density, potential and electric field, electron temperature, and other related parameters. To achieve this simulation, we decided to include various gases as additives to the nitrogen gas. This process involves a detailed examination of the reactions between these gases and electrons with the species present in the discharge phenomenon.

**Electrical discharge simulation between two electrodes**

To achieve the objectives of the present investigation, the dielectric barrier discharge simulation was carefully performed at 1 atmosphere pressure and 400 Kelvin temperature. Our goal is to create an accurate simulation of electrical discharge in the presence of oxygen and nitrogen gases, which are known for their reactive oxygen and nitrogen species (RONS). In this regard, we initially used pure nitrogen gas ($N_2$) and then considered the use of air, which is a mixture of nitrogen and oxygen. These gases were placed between two parallel plates, and then, by applying a sinusoidal alternating voltage, the necessary conditions for plasma generation were established. This process not only leads to plasma production, but also accurately displays the physical behaviors and chemical reactions in various environments under simulated conditions. Although the process of simulating electrical discharge in two dimensions may be practical and feasible, some challenges and issues arising from this approach lead us to choose an alternative. One of the main problems is the increased simulation time and the greater complexity and difficulty in interpreting and analyzing the results. For this reason, our preference and interest shift towards using a one-dimensional simulation, as this method can be more efficient. To perform the desired electrical discharge simulation, we used version 6.1 of the COMSOL software. In this program, we used the Plasma module, which is specifically designed for modeling plasma phenomena. Additionally, for greater accuracy and more detailed analysis, we utilized the Plasma- Plas submodule. After selecting a sinusoidal voltage for the power electrode, it is necessary to correctly define and input the time-dependent behavior of the electrical discharge process. To perform this, in the software, we navigated to the study section and selected the Time Dependent option.

**Geometry and Meshing**

The geometry used in this study is one-dimensional (line and points). The COMSOL software employs the Finite Element Method (FEM) strategy. Figure 1 illustrates the geometry and boundary conditions of the air-gas discharge system using a dielectric barrier.

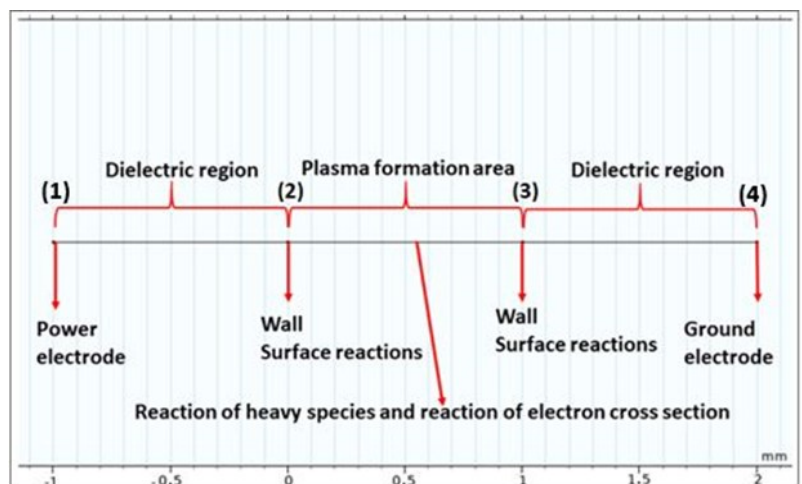

Figure 1. Schematic of the system geometry and boundary conditions

The geometry of the system includes two electrodes (power and ground), two dielectrics, the plasma formation region, and the surface reactions at the walls. The starting and ending points of our design clearly represent the two electrodes (1 and 4), whereas the distance between these electrodes, along with their opposing points, forms our dielectric (between 1 and 2, between 3 and 4). The intermediate space, located between the two dielectrics, is referred to as the gap region, where all the essential and vital reactions in this system take place (between 2 and 3). The distance between the two electrodes is precisely set at 3 mm, while the gap length is determined to be 1 mm. The dielectrics used in this design are 1 mm long and have a considerable thickness of 1 mm. Figure 2 presents a two-dimensional schematic of the geometry of the system.

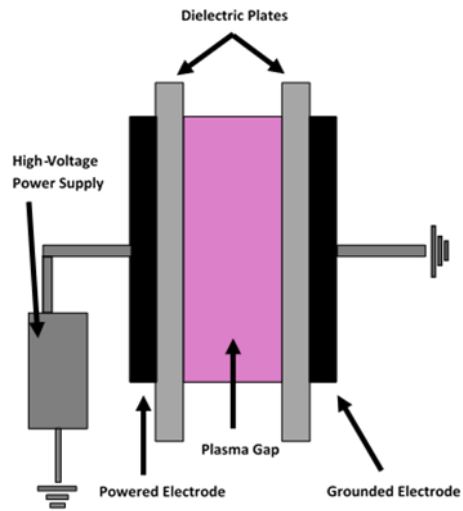

Figure 2. Two-dimensional diagram of the system geometry.

The point-based meshing (networking) used in simulation is discussed in this study. The COMSOL software uses FEM to decompose complex problems into smaller, more solvable components. To achieve balance, we use a variable-density meshing approach. In areas where events are of greater significance, we use finer meshing to model more details (Minimum element size=6E-4 [mm]). Conversely, in more distant and less critical regions, we use larger elements (Maximum element size= 3E-2 [mm]), which help reduce the overall computational load. Figure 3 shows an overall view of the meshing used in this study, which appears uniform due to the large number of elements. Moreover, Table 1 presents the general features of the meshing.

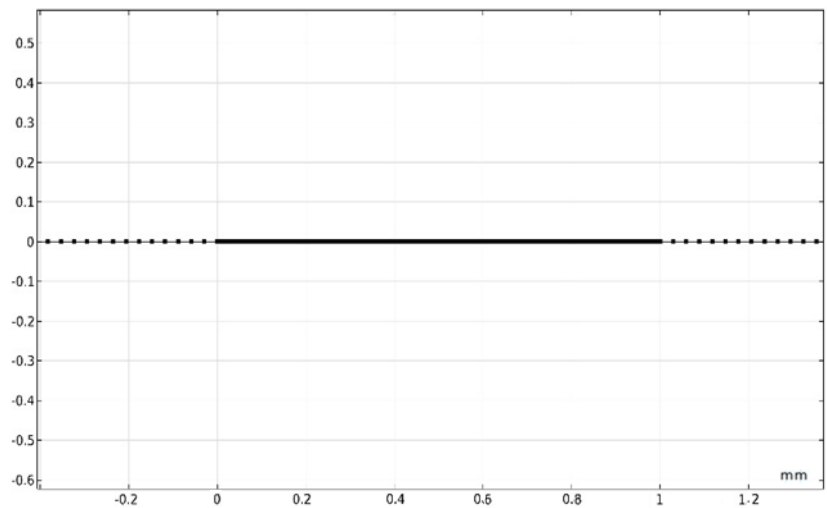

Figure 3. Schematic of the meshing of the system.

Table 1. General features of the meshing.

| Description | Value |
|---|---|
| Maximum element size | 3E-2 [mm] |
| Minimum element size | 6E-4 [mm] |
| Curvature factor | 0.2 |
| Predefined size | Extremely fine |

**Reactions and Parameters**

Air is made up of 78% nitrogen and 21.5% oxygen, with the rest being various gases, including moisture, which we consider to be air vapor (0.5 %). In this study, 87 reactions between heavy species and electron reactions with neutral species in the plasma region, as well as other surface reactions for secondary electron emission and plasma stability in the wall region, were defined. In line with the conducted research, nitrogen gas was first introduced into the gap region, followed by air gas. Under the specific conditions present, these gases undergo various electrochemical and physical reactions in the region of interest.

Table 2 clearly shows the electron impact reactions in detail for the species present in the nitrogen gas discharge state. This table represents electron impact reactions with nitrogen molecules, $N_2s$ (excited nitrogen), and $N_2^+$ (one-ionized nitrogen).

Table 2. Electron impact reactions [18], [22], [23].

| Reaction | Formula | Type | $\Delta E(ev)$ |
|---|---|---|---|
| 1 | $e + N_2 \rightarrow e + N_2$ | Elastic | - |
| 2 | $e + N_2 \rightarrow e + N_2s$ | Excitation | 0.02 |
| 3 | $e + N_2 \rightarrow e + N_2s$ | Excitation | 0.29 |
| 4 | $e + N_2 \rightarrow e + N_2s$ | Excitation | 0.291 |
| 5 | $e + N_2 \rightarrow e + N_2s$ | Excitation | 0.59 |
| 6 | $e + N_2 \rightarrow e + N_2s$ | Excitation | 0.88 |
| 7 | $e + N_2 \rightarrow e + N_2s$ | Excitation | 1.17 |
| 8 | $e + N_2 \rightarrow e + N_2s$ | Excitation | 1.47 |
| 9 | $e + N_2 \rightarrow e + N_2s$ | Excitation | 1.76 |
| 10 | $e + N_2 \rightarrow e + N_2s$ | Excitation | 2.06 |
| 11 | $e + N_2 \rightarrow e + N_2s$ | Excitation | 2.35 |
| 12 | $e + N_2 \rightarrow e + N_2s$ | Excitation | 6.17 |
| 13 | $e + N_2 \rightarrow e + N_2s$ | Excitation | 7 |
| 14 | $e + N_2 \rightarrow e + N_2s$ | Excitation | 7.35 |
| 15 | $e + N_2 \rightarrow e + N_2s$ | Excitation | 7.8 |
| 16 | $e + N_2 \rightarrow e + N_2s$ | Excitation | 8.4 |
| 17 | $e + N_2 \rightarrow e + N_2s$ | Excitation | 8.16 |
| 18 | $e + N_2 \rightarrow e + N_2s$ | Excitation | 8.55 |
| 19 | $e + N_2 \rightarrow e + N_2s$ | Excitation | 8.89 |
| 20 | $e + N_2 \rightarrow e + N_2s$ | Excitation | 11.03 |
| 21 | $e + N_2 \rightarrow e + N_2s$ | Excitation | 11.88 |
| 22 | $e + N_2 \rightarrow e + N_2s$ | Excitation | 12.25 |
| 23 | $e + N_2 \rightarrow e + N_2s$ | Excitation | 13 |
| 24 | $e + N_2 \rightarrow 2e + N_2^+$ | Ionization | 15.6 |
| 25 | $e + N_2 \rightarrow e + N_2a3s$ | Excitation | 6.72 |
| 26 | $e + N_2a3s \rightarrow e + N_2$ | Excitation | -6.72 |
| 27 | $e + N_2 \rightarrow e + N + N$ | Excitation | 8 |
| 28 | $e + H_2O \rightarrow e + H + OH$ | Excitation | 7.1 |

By inputting the electron collision cross-section data and its energy in each of the aforementioned states, which were provided as inputs to the software, accurate values of species density and other important information will be obtained. Other reactions occurring in the nitrogen gas discharge $N_2$ in the specified gap are shown in Table 3. For these reactions, the software receives a rate constant, denoted as $K^f$, as input, and these reactions are considered irreversible.

Table 3. Other gap space reactions for nitrogen gas discharge [18].

| Reaction | Formula | Type | Rate Constant $K^f(\frac{m^3}{s.mol})$ |
|---|---|---|---|
| 1 | $e + N^+ \rightarrow N$ | Recombination | $3.5 \times 10^{-18}$ |
| 2 | $e + N_2 \rightarrow 2e + N + N^+$ | Dissociative ionization | $2.4 \times 10^{-23}$ |
| 3 | $e + N_2 \rightarrow 2N$ | Dissociative | $2 \times 10^{-17}$ |
| 4 | $e + N_2^+ \rightarrow 2N$ | - | $2.8 \times 10^{-13}$ |
| 5 | $N^+ + N_2 \rightarrow N + N_2^+$ | Charge exchange | $10^{-17}$ |

In the Table 3, $N^+$ denotes a singly ionized nitrogen atom, $N_2^+$ denotes a singly ionized nitrogen molecule, and N denotes a nitrogen atom.

Additionally, other reactions known as surface reactions are defined at the wall boundary (boundaries 2 and 3 in figure 1). In these reactions, the ionized and excited species combine with electrons and return to their neutral state. Table 4 presents the essential surface reactions for simulating the electric discharge of nitrogen gas in a comprehensive and accurate manner.

Table 4. Surface reactions required in air discharge simulation [18].

| Reaction | Formula | Sticking coefficient |
|---|---|---|
| 1 | $O_2s \rightarrow O_2$ | 1 |
| 2 | $O_2a1d \rightarrow O_2$ | 1 |
| 3 | $O_2b1s \rightarrow O_2$ | 1 |
| 4 | $O_245 \rightarrow O_2$ | 1 |
| 5 | $O^- \rightarrow O$ | 1 |
| 6 | $O1d \rightarrow O$ | 1 |
| 7 | $O1s \rightarrow O$ | 1 |
| 8 | $N_2^+ \rightarrow N_2$ | 1 |
| 9 | $N_2a3s \rightarrow N_2$ | 1 |

To simulate the electric discharge in air, we require more complex processes that involve additional reactions. These reactions include the effect of electrons on oxygen species as well as interactions between nitrogen, oxygen, and water vapor species. The oxygen species, due to their high reactivity, are considered one of the key factors in this system. Table 5 shows the electron impact reactions on oxygen molecules for the electric discharge simulation in air.

Table 5. Electron impact reactions on oxygen molecules [22], [23].

| Reaction | Formula | Type | $\Delta E(ev)$ |
|---|---|---|---|
| 1 | $e + O_2 \rightarrow e + O_2$ | Elastic | - |
| 2 | $e + O_2 \rightarrow O + O^-$ | Attachment | - |
| 3 | $e + O_2 \rightarrow e + O_2$ | Excitation | 0.02 |
| 4 | $e + O_2 \rightarrow e + O_2$ | Excitation | 0.19 |
| 5 | $e + O_2 \rightarrow e + O_2$ | Excitation | 0.19 |
| 6 | $e + O_2 \rightarrow e + O_2$ | Excitation | 0.38 |
| 7 | $e + O_2 \rightarrow e + O_2$ | Excitation | 0.38 |
| 8 | $e + O_2 \rightarrow e + O_2$ | Excitation | 0.57 |
| 9 | $e + O_2 \rightarrow e + O_2$ | Excitation | 0.75 |
| 10 | $e + O_2 \rightarrow e + O_2a1d$ | Excitation | 0.977 |
| 11 | $e + O_2a1d \rightarrow e + O_2$ | Excitation | -0.977 |
| 12 | $e + O_2 \rightarrow e + O_2b1s$ | Excitation | 1.627 |
| 13 | $e + O_2b1s \rightarrow e + O_2$ | Excitation | -1.627 |
| 14 | $e + O_2 \rightarrow e + O_245$ | Excitation | 4.5 |
| 15 | $e + O_245 \rightarrow e + O_2$ | Excitation | -4.5 |
| 16 | $e + O_2 \rightarrow e + O + O$ | Dissociative | 6 |
| 17 | $e + O_2 \rightarrow e + O + O1d$ | Excitation | 8.4 |
| 18 | $e + O_2 \rightarrow e + O1s$ | Excitation | 9.97 |
| 19 | $e + O_2 \rightarrow 2e + O_2^+$ | Excitation | 12.06 |

Next, in Table 6, other reactions required to simulate the electrical discharge of air gas between two dielectrics are presented:

Table 6. Other reactions required in the simulation of electrical discharge in air [18], [19].

| Reaction | Formula | Rate Constant $K^f(\frac{m^3}{s.mol})$ |
|---|---|---|
| 1 | $O + O_2 + O_2 \rightarrow O_3 + O_2$ | $6 \times 10^{-46} \times (1.3^{-2.8})$ |
| 2 | $O + O_2 + N_2 \rightarrow O_3 + N_2$ | $5.6 \times 10^{-46} \times (1.3^{-2.8})$ |
| 3 | $O + O_3 \rightarrow O_2 + O_2$ | $8 \times 10^{-18} \times exp(^{-2060}/_{40})$ |
| 4 | $N + O_3 \rightarrow NO + O_2$ | $1 \times 10^{-22}$ |
| 5 | $N_2^+ + N_2 + O_2 \rightarrow N_4^+ + O_2$ | $18 \times 10^6$ |
| 6 | $N_4^+ + O_2 \rightarrow O_2^+ + N_2 + N_2$ | $15 \times 10^7$ |
| 7 | $N + O_2 \rightarrow NO + O$ | $1.5 \times 10^{-7} \times exp(8)$ |
| 8 | $N + NO \rightarrow N_2 + O$ | $2.1 \times 10^{-11} \times exp(0.25)$ |
| 9 | $2N + NO \rightarrow N_2 a3s + NO$ | $6.1 \times 10^2$ |
| 10 | $O1s + NO \rightarrow O1d + NO$ | $2.2 \times 10^8$ |
| 11 | $NO + OH + O_2 \rightarrow HNO_2 + O_2$ | $2.2 \times 10^5$ |
| 12 | $NO + OH + N_2 \rightarrow HNO_2 + O_2$ | $2.2 \times 10^5$ |
| 13 | $HNO_2 + OH \rightarrow NO_2 + H_2O$ | $7.8 \times 10^4$ |
| 14 | $NO_2 + H \rightarrow NO + OH$ | $7.7 \times 10^7$ |
| 15 | $H_2O + O \rightarrow OH + OH$ | $9.6 \times 10^5$ |
| 16 | $OH + OH \rightarrow H_2O + O$ | $1.1 \times 10^6$ |
| 17 | $O + OH \rightarrow O_2 + H$ | $2.3 \times 10^{-17} \times exp(0.3)$ |
| 18 | $OH + O_3 \rightarrow HO_2 + O_2$ | $4 \times 10^4$ |
| 19 | $OH + HO_2 \rightarrow H_2O + O_2$ | $4.7 \times 10^7$ |
| 20 | $HO_2 + O \rightarrow OH + O_2$ | $9 \times 10^6$ |
| 21 | $HO_2 + O_3 \rightarrow OH + O_2 + O_2$ | $1.2 \times 10^3$ |
| 22 | $H + O_2 + O_2 \rightarrow HO_2 + O_2$ | $1.9 \times 10^4$ |
| 23 | $H + O_3 \rightarrow OH + O_2$ | $1.2 \times 10^6$ |
| 24 | $H + O_3 \rightarrow HO_2 + O$ | $1.2 \times 10^6$ |
| 25 | $H + HO_2 \rightarrow OH + O$ | $10^7$ |
| 26 | $H + OH \rightarrow NO + H$ | $1.7 \times 10^7$ |

In Table 6, $O_2$ represents an oxygen molecule, $O_3$ represents an ozone molecule, O represents an oxygen atom, $O_2^+$ represents a singly positively charged oxygen molecule, $O_2^-$ represents a singly negatively charged oxygen molecule, $H_2O$ represents water vapor, NOx compounds (NO, $NO_2$, $NO_3$), and $N^+$ represents a singly positively charged nitrogen atom.

**Initial Parameters**

The measurement system in this simulation is SI. The simulation was carried out at three frequencies: 500 Hz, 50 kHz, and 500 kHz, with an applied peak-to-peak voltage of 2 kV. The diameter of the plate was set at 0.1 meters. The initial parameters which show in Table 7 were entered in the Initial Values section of the Plasma module. The time step was chosen as $1 \times 10^{-5}$ for 500 Hz, $1 \times 10^{-7}$ for 50 kHz, and $1 \times 10^{-8}$ for 500 kHz.

Table 7. Initial parameters of simulation

| Initial parameters | Parameters value |
|---|---|
| Pressure | 760 [Torr] |
| Temperature | 400 [k] |
| Initial mean electron energy | 5 [V] |
| Initial Electron density | $10^{13}$ [1/m$^3$] |
| Electron mobility | Calculation from the electron collision cross section |

## Physics and Equations

Now, we add surface reactions to the model; these reactions specifically describe the neutralization of ions on the electrodes. To ensure the stability of the electrical discharge, the presence of secondary electron emission is essential. Therefore, it is necessary to specify and input the emission coefficient and an estimate of the average energy of secondary electrons based on the ionization threshold energy and the metal surface's work function. For these reactions to make sense, a wall must be defined; therefore, boundaries 2 and 3, as shown in Figure 1, will act as the wall. However, an important point in the simulation process is that plasma-related problems must be solved under electrically neutral initial conditions. To this end, in the nitrogen ion species section (Species: N$_2$), the option "Initial Value from Electroneutrality Constraint" is enabled for both electrical discharge simulation scenarios.

Part of the equations includes the drift-diffusion equations solved for each species, such as electrons, positive ions, and negative ions. Equations 1 and 2 represent the electron transport and the sources of electron production and loss, respectively.

$$\frac{\partial n_e}{\partial t} + \nabla.(-\overrightarrow{D_e}\nabla n_e - \overrightarrow{\mu_e}\overrightarrow{E}n_e) - R_e = 0 \qquad (1)$$

$$R_e = \sum_{j=1}^{M} x_j \, \alpha_j N_n |\Gamma_e| \qquad (2)$$

Where, in Equation (1), the electron density, mobility and diffusion coefficient are denoted as n$_e$, μ$_e$ and D$_e$ respectively. The electric field vector and the electron production and loss source term are represented by $\overrightarrow{E}$ and R$_e$, respectively. Equation (3), known as the Poisson equation, represents the electric field distribution in the region between two electrodes also calculates the electrostatic field:

$$\nabla.(\varepsilon_0 \varepsilon_r \overrightarrow{E}) = \rho \qquad (3)$$

Where ε$_0$, ε$_r$ and ρ represent the permittivity of the free space, the relative dielectric constant, and the net charge density of the plasma species, respectively. Equations (4) and (5) represent the secondary emission flux resulting from the collisions of positive and negative ions with the electrode surface, leading to the ejection of electrons from the electrode surface.

$$\overrightarrow{\Gamma_e} = \gamma n_n \mu_n |\overrightarrow{E}| \qquad (4)$$

$$\overrightarrow{\Gamma_e} = \gamma n_P \mu_P |\overrightarrow{E}| \qquad (5)$$

In which γ represents the secondary emission coefficient.

The electron energy density is represented in Equation (6), and the energy rate due to inelastic collisions is given in Equation (7).

$$\frac{\partial n_\varepsilon}{\partial t} + \nabla.\overrightarrow{\Gamma_\varepsilon} + \overrightarrow{E}.\overrightarrow{\Gamma_\varepsilon} = R_\varepsilon - (\vec{u}.\nabla)n_\varepsilon \qquad (6)$$

$$R_\varepsilon = S_{en} + \frac{Q + Q_{gen}}{q} \tag{7}$$

In Equation (7), $S_{en}$ represents the collision power loss ($\frac{W}{m^3}$), Q is the electron heat source (typically due to the magnetic field), $Q_{gen}$ is the main heat source, and q is the electric charge (C). Surface charge accumulation at the boundary of the walls (boundaries 2 and 3):

$$\frac{\partial \sigma_s}{\partial t} = n.J_i + n.j_e \tag{8}$$

Where $\sigma_s$, $J_i$ and $j_e$ are the surface charge density, ionic flux density and electric flux density, respectively. To calculate $\sigma_s$, we have:

$$\sigma_s = -n.(D1 - D2) \tag{9}$$

$D1 - D2$ are displacement fields across an interface, multiplied by the surface normal (n).

The continuity equation (10) for ion species is as follows:

$$\frac{\partial n_i}{\partial t} + \nabla.(n_i \vec{u}) = -\nabla.(\mu_i n_i q_i \nabla \varphi - D_i \nabla n_i) + S_i \tag{10}$$

In total, in the equations mentioned above, $\vec{\Gamma}_e$ is the electron flux, $R_\varepsilon$ is the energy rate of inelastic collisions, $n_\varepsilon$ is the electron energy density, $S_i$ is the ion production source, $n_i$ is the ion density and $\varphi$ is the electrostatic potential. According to Figure 1, for equations (1), (2), (3), (6), (7), and (10), all occur in the boundaries between 2 and 3, in the plasma region. And equation (4) and (5) are used in boundary 2 and 3, which secondary emission will happen. Meanwhile, the boundary conditions of the dielectric region, namely the region between boundaries 1 and 2 and the region between boundaries 3 and 4, were defined as charge conservation.

## III. SIMULATION RESULTS AND ANALYSIS

**Mesh-independent curve**

According to the curve plotted in Figure 4, changes in the mesh elements show little variation in the electric current values at a frequency of 50 kHz and a peak-to-peak voltage of 2 kV.

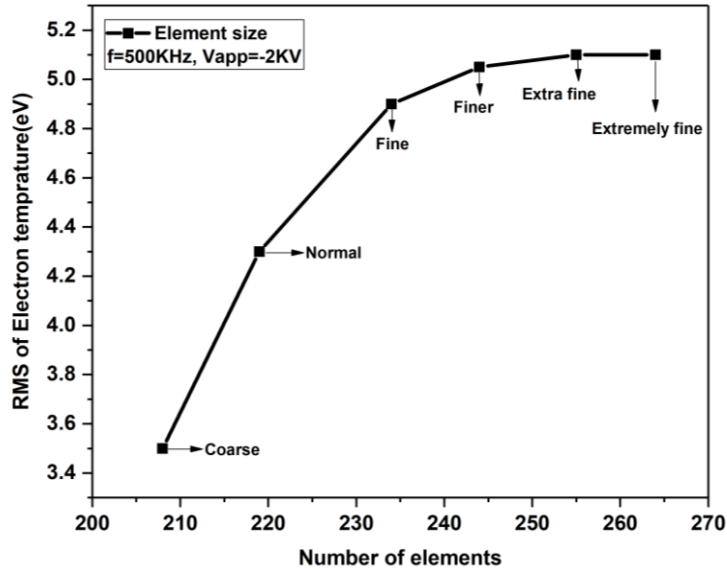

Figure 4. The mesh-independent curve of the RMS of electron temperature at the center of the plasma discharge gap

As can be seen, no changes in the electron temperature values are observed from the default element size of Fine onward. Therefore, in the present study, the Extremely Fine setting was considered by default.

Figure 5 shows the voltage-current waveform at three frequency levels: 500 Hz, 50 kHz, and 500 kHz, with a peak-to-peak voltage of 2 kV. In the voltage-current curve the current will change when we use the sinusoidal voltage, according to the collision rate and the time dependence of the plasma equations. At higher frequencies, the plasma experiences rapid charging and discharging of the dielectric barrier, which may lead to unequal current amplitudes and asymmetric waveform shapes during the early cycles. These effects are especially visible during the ignition phase, where the space charge and wall charge distributions are still developing. To have better estimation of concentration of species, we use root mean square (RMS) of two periods in this simulation.

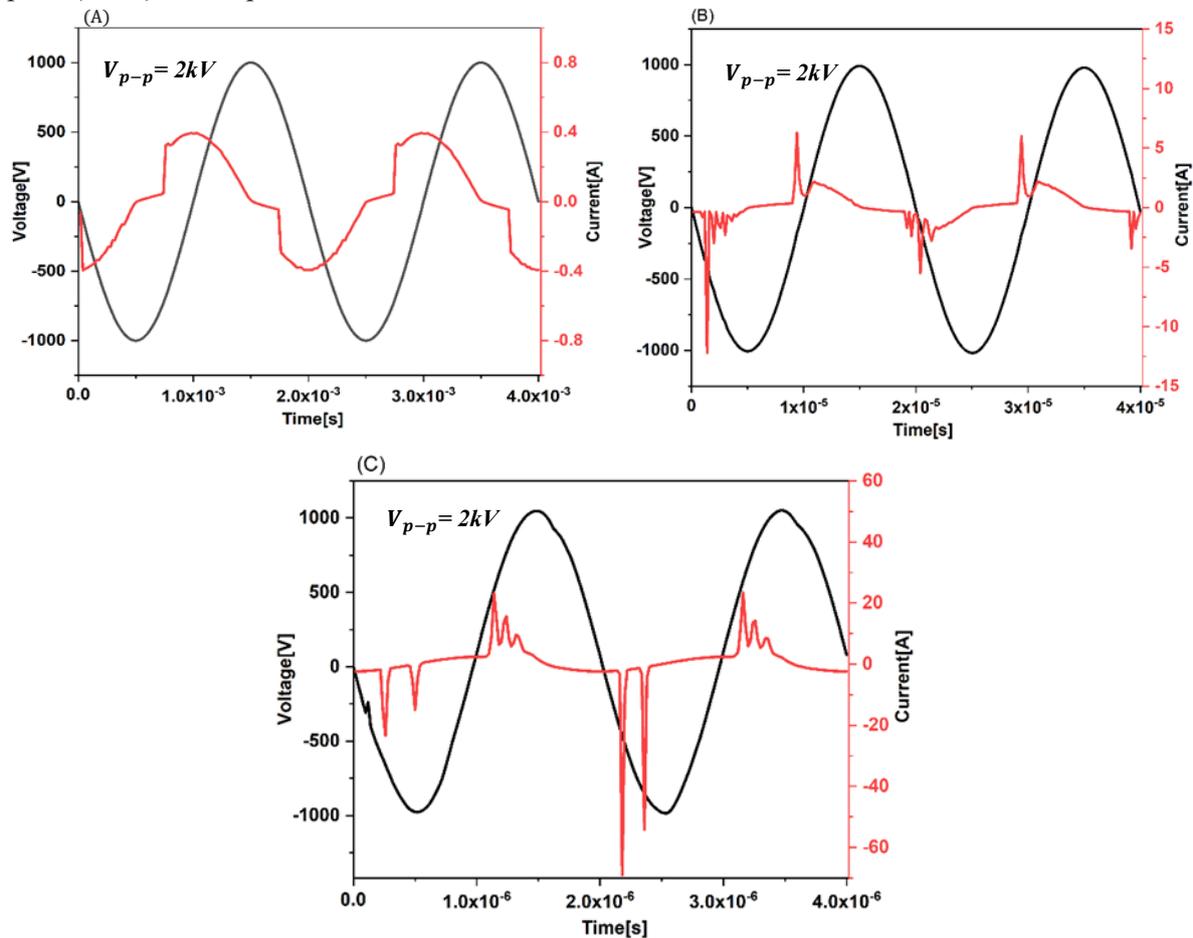

Figure 5. The V-I characteristic curves at three frequency levels: A) 500 Hz, B) 50 kHz and C) 500 kHz.

Figures (6) to (8) show the logarithmic RMS curves of different distributions at three frequency levels in the plasma region. The reason for plotting the logarithmic curve was that the differences between the curves at certain frequencies were large, and the curve lines either overlapped or were spaced far apart. Another reason was the inability to observe the details. As shown in figure 6 higher frequency shows higher temperature and electron density which are 5.5 eV and around $10^{18}$ (1/m³) in maximum, respectively.

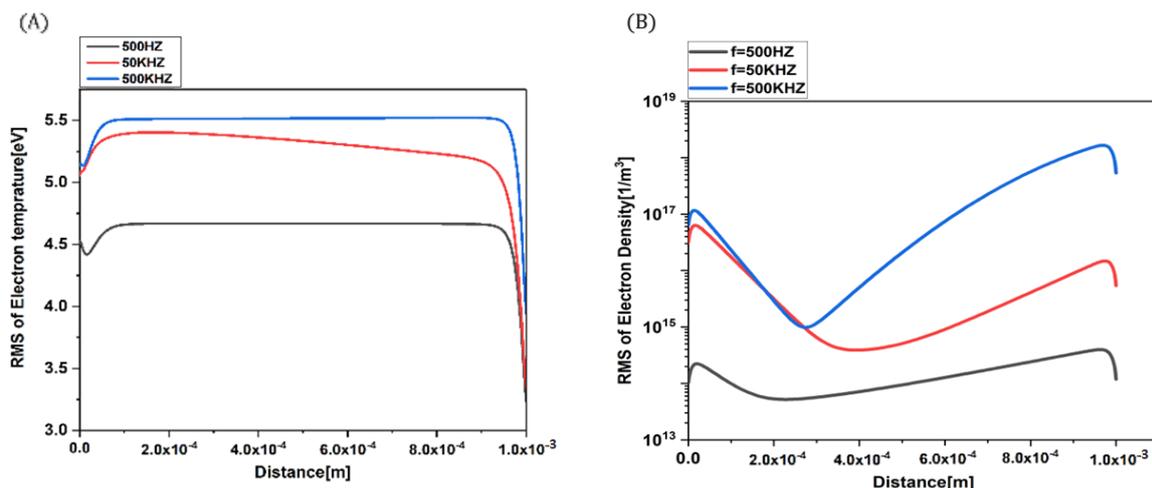

Figure 6. Logarithmic RMS curves of the A) electron temperature and (B) electron density distributions at three frequency levels in the plasma region.

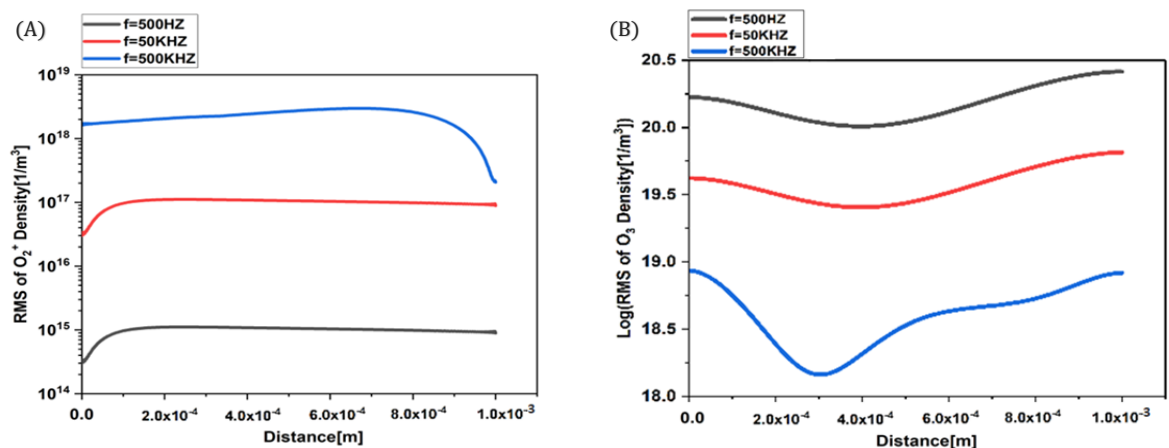

Figure 7. The RMS curve of the A) O₂+ ion density distribution and the B) O₃ species at three frequency levels in the plasma region.

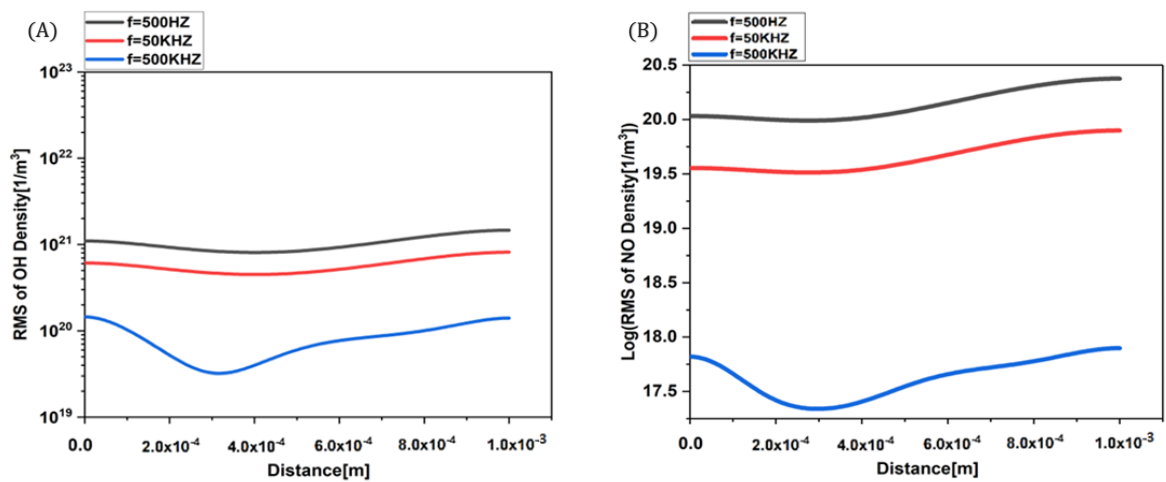

Figure 8. The logarithmic RMS curve of the A) OH and B) NO species density distribution at three frequency levels in the plasma region.

Figure 7 and 8 show concentration of different species with different frequency in plasma reactor, as shown in figures increasing frequency can decreasing density of NO, OH and O₃, but density of O₂⁺ shows an increasing trend toward increasing frequency. At higher frequencies, the electron energy distribution function (EEDF) changes, and ion recombination rates can also change [24]. At higher frequencies, the electric field changes more rapidly, which can result in more energetic electron collisions, increasing the formation of $O_2^+$ through ionization of $O_2$ molecules. Reduced densities of other reactive species might indicate that recombination or secondary reactions are more dominant at higher frequencies, or that the plasma chemistry is shifting towards stronger ionization rather than dissociation.

**Two-dimensional voltage distribution**

The two-dimensional voltage distribution diagrams at three frequency levels are shown if Figure 9.

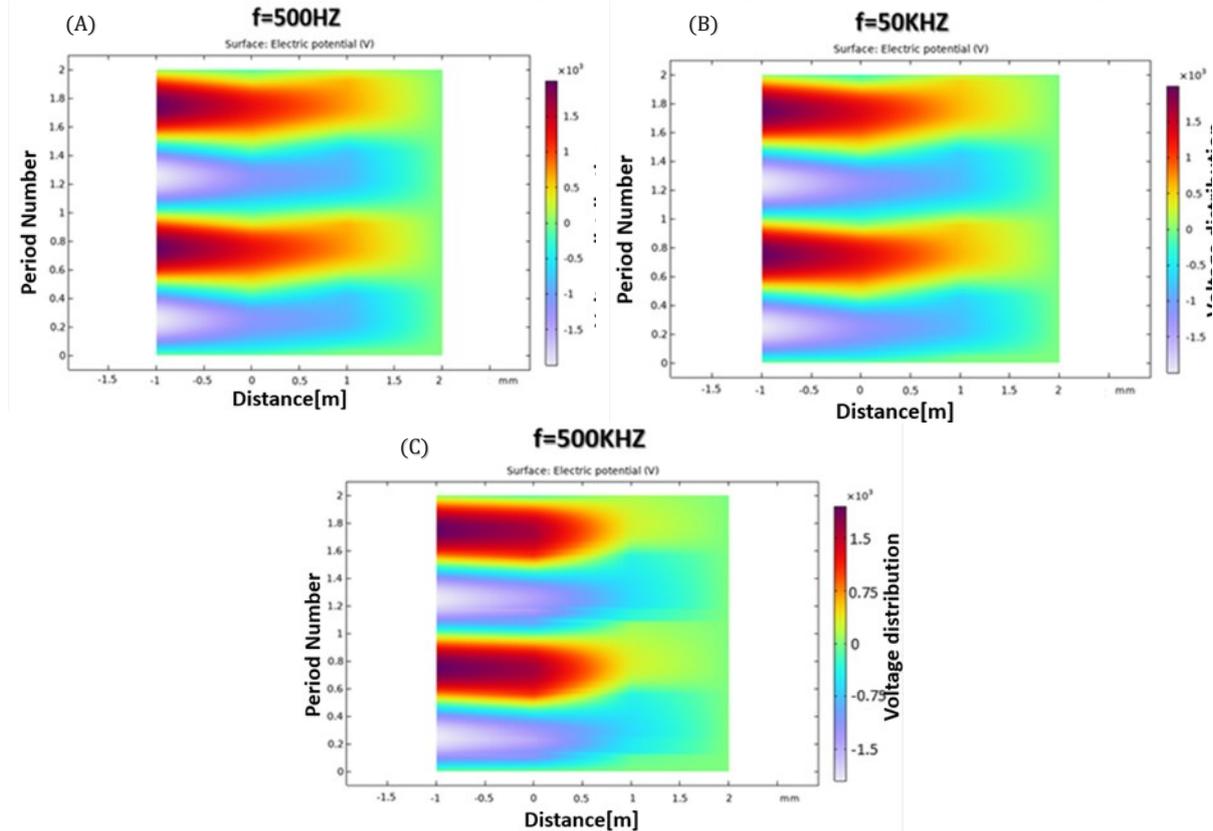

Figure 9. The two-dimensional voltage distribution diagram at three frequency levels: A) 500 Hz, B) 50 kHz, and C) 500 kHz.

In Figure 9, the blue distribution represents the ground electrode, and the red distribution represents the power electrode. The horizontal axis shows the distance, the left vertical axis indicates the number of cycles, and the right vertical axis represents the voltage distribution.

**Electric field Distribution**

In Figure 10, the two-dimensional electric field shape illustrates that the field in the gap between the two electrodes is weaker than in the dielectric region, and this is due to the accumulation of surface charge, which tends to shield the electric field.

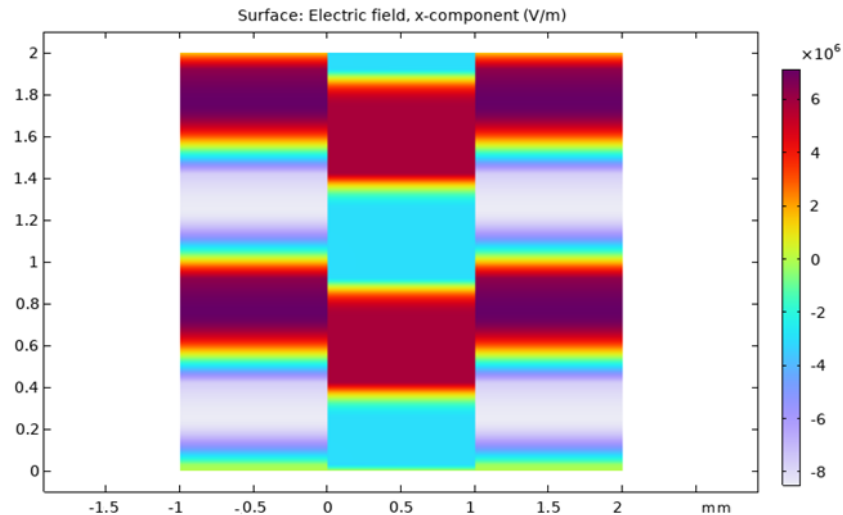

Figure 10. Electric field distribution at a frequency of 50 kHz

From a distance of -1 to 0, the electric field is within the dielectric range. The distance from 0 to 1 is the region where the plasma forms. Similarly, the distance from 1 to 2 is within the dielectric electric field. It should be noted that the dielectric region appears darker, whereas the plasma region is lighter. The reason for the breakdown of the electric field is the accumulation of surface charge on the dielectric surface. Another physical reason is that the field inside the dielectric is stronger than the field inside the plasma.

The quantity acquired for the species was contrasted with earlier research. Because the plasmas used in the previous research were specially designed for different objectives, the plasma's properties, such voltage, were set at higher levels, which increased the discharge gap. Nonetheless, the density of particles is precisely proportional to the air atmospheric pressure plasma [25]–[27]. The electron density obtained from our COMSOL simulation of DBD plasma is in close agreement with the values reported in [28]. In this overview, different methods used in plasma electron density measurement had been investigated. And they reported the electron density range $10^9$-$10^{18}$ (1/cm$^3$) which result obtained in this worked showed the same range [18], [29]. The electron temperature in simulation is 5.5 in maximum which is in the range. The electron temperature for air gas discharge is 2-10 eV depending on the input parameter [30], [31]. Effect of frequency on O3 generation efficiency had been investigated by Huang et al.[32]. The results indicate that ozone generation efficiency rises with increasing ozone concentration at 50 Hz: conversely, this efficiency declines with higher ozone concentration at 2.6 kHz and 20 kHz. Similarly, in reference [33] showed a one-dimensional fluid model including 17 species and 65 reactions is used to investigate the effects of driving frequency on the generation and elimination of ROS in atmospheric radio frequency helium-oxygen discharges. The computational findings indicate that when frequency increases, the densities of ROS (atomic oxygen, ozone, excited atomic oxygen, singlet delta oxygen) consistently drop at a fixed power density. Also, the study of the effect of frequency on temperature and electron density at radio frequencies (100 MHz) up to 2.45 GHz has been explained by Kwon et al. [34]. Moreover, importance of pulse repetition in air discharge had been investigated by Kushner et al. [35]. Although they used a numerical simulation with different parameters in comparison current work, the effect of frequency on RON's production is obvious.

The results show only minor discrepancies, which can be attributed to differences in model assumptions, boundary conditions, or numerical methods. This validation confirms the reliability of our simulation approach in accurately capturing the plasma behavior.

## IV. CONCLUSION

In this study, the electron temperature profile characteristic of this simulation was obtained and found to match the electron temperature profile derived from the experimental data in previous articles. The results of the study demonstrate that the frequency of the applied voltage significantly influences the density of species generated in the atmospheric pressure gas discharge. Considering the impact of each species in various plasma applications, adjusting the frequency of the applied voltage can optimize and achieve the desired values under different conditions. These findings indicate that the density of ionized oxygen and nitrogen species increases with increasing applied frequency. Furthermore, the results showed that increasing the frequency of the applied voltage in gas discharge enables plasma generation at lower voltages while also altering the density of influential species such as ozone density, oxygen atoms, and NOx species. Moreover, even with identical power consumption and applied voltage at different frequencies, observable variations are still present.

**Funding**

Funded by the EU NextGenerationEU through the Recovery and Resilience Plan for Slovakia under project No. 09I03-03-V03-00033 EnvAdwice and Slovak Research and Development Agency APVV-22-0247.

**Conflict of interest**

The authors declare no competing interests in relation to this study.